\documentstyle[12pt]{article}
\newcommand{\be}{\begin{equation}}
\newcommand{\ee}{\end{equation}}
\newcommand{\bea}{\begin{eqnarray}}
\newcommand{\eea}{\end{eqnarray}}

\newfont{\bg}{cmr10 scaled\magstep4}

\newcommand{\bigzerou}{\smash{\lower1.7ex\hbox{\bg 0}}}
\begin{document}
\title{
De Rham Cohomology of $SO(n)$ by Supersymmetric Quantum Mechanics}
\author{Kazuto Oshima \\ \\
\sl Gunma College of Technology,Maebashi 371,Japan \\
\sl e-mail:oshima@nat.gunma-ct.ac.jp}
\date{ }
\maketitle
\abstract{We give an elementary derivation of the de Rham cohomology of 
$SO(n)$ in terms of Supersymmetric Quantum Mechanics. Our analysis is based
on  Witten's Morse theory. We show reflection symmetries of the theory
are useful to select true vacuums. The number of the selected  vacuums
will agree with the de Rham cohomology of $SO(n)$.}
\maketitle
\\ 
\baselineskip=24pt
\section{Introduction}
In his pioneering work in 1982[1], Witten proposed a physical interpretation
of de Rham cohomology. He considered  supersymmetric quantum mechanics
on a manifold with a potential derived from a Morse function $h$. His idea
is to identify the theory with de Rham theory. For each critical point of $h$
one classical vacuum exists. We can select true vacuums by examining 
instanton
effects between adjacent classical vacuums. The number of classical vacuums
is not smaller than that of true vacuums, which explains the Morse
inequalities.

To our knowledge his program has been carried out only for few examples.
The group manifold $SO(n)$ is interesting, because it has
many couples of adjacent critical points. Some years ago, Yasui et al.[2] 
investigated $SO(3)$. Recently, the author has tackled $SO(4)$[3].
Owing to reflection symmetries of the theory
a pair of instanton effects cancel each other in some processes.
The purpose of this letter is to give a generalization to $SO(n)$.
We give an elementary derivation of the de Rham cohomology of $SO(n)$
based on an instanton picture.

In Section 2 we introduce  supersymmetric quantum mechanics on $SO(n)$.
In Section 3 we exhibit classical vacuums of the theory. In section 4
we explain reflection symmetries for $SO(n)$. In Section 5 we state how
to select true vacuums. In the last section we treat $SO(5)$ as an example.

\section{Supersymmetric Quantum Mechanics on $SO(n)$}
The  supersymmetric hamiltonian on a manifold is given by 
\begin{equation}
\hat{H}=-{1 \over 2}(d_{h} d_{h}^{\dagger} +d_{h}^{\dagger} d_{h}),
\end{equation}
where $d_{h}=e^{-h}de^{h}, d_{h}^{\dagger}=e^{h}d^{\dagger}e^{-h},
d$ is the exterior derivative and $d^{\dagger}$ is its adjoint operator.
The exterior multiplication $e_{dx ^{\mu}}$ and the interior multiplication
$i_{\partial \over \partial x^{\mu}}$ can be identified with the fermion 
creation
operator $\hat{\psi}^{* \mu}$ and the annihilation operator 
$\hat{\psi} _{\mu}$ and we have
\begin{equation}
d=\hat{\psi}^{* \mu} \nabla _{\mu}, \quad  
d^{\dagger}=g^{-{1 \over 2}}\nabla _{\mu} g^{1 \over 2}g^{\mu \nu}\hat{\psi} 
_{\nu},
\end{equation}
where $\nabla _{\mu}$ is the covariant derivative
\begin{equation}
\nabla _{\mu}={\partial \over \partial x^{\mu}}
-\Gamma_{\mu \nu}^{\lambda}\hat{\psi} ^{* \nu}\hat{\psi}_{\lambda}. 
\end{equation}
The hamiltonian $\hat{H}$ (1) takes the form:
\begin{equation}
2\hat{H} =-g^{-{1 \over 2}} \nabla _{\mu}g^{1 \over 2}g^{\mu \nu} \nabla 
_{\nu}
+R_{\mu \nu \sigma \tau}
\hat{\psi}^{\sigma}\hat{\psi}^{* \tau}\hat{\psi}^{* \nu}\hat{\psi} ^{\mu}
+g^{\mu \nu} {\partial h \over \partial x^{\mu}}{\partial h \over \partial 
x^{\nu}}
+H_{\mu \nu}[\hat{\psi} ^{* \mu},\hat{\psi}^{\nu}],
\end{equation}
where $g_{\mu \nu}$ and $R_{\mu \nu \sigma \tau}$ are the Riemann metric and 
tensor, and
 $H_{\mu \nu}$ is the Hessian matrix
\begin{equation}
H_{\mu \nu}=(\partial_{\mu}\partial_{\nu}-\Gamma_{\mu 
\nu}^{\lambda}\partial_{\lambda})h.
\end{equation}
Corresponding Laglangian is 
\bea
{\cal L} ={1 \over 2}g_{\mu \nu}{dx^{\mu} \over dt}{dx^{\nu} \over dt}
+{1 \over 2}g^{\mu \nu}{\partial h \over \partial x^{\mu}}
{\partial h \over \partial x^{\nu}}
+\psi ^{* \mu}({d \over dt}\psi_{\mu}-\Gamma_{\mu 
\nu}^{\lambda}\psi_{\lambda}
{dx^{\nu} \over dt}) \nonumber  \\ +H_{\mu \nu}\psi ^{* \nu}\psi^{\mu} 
+{1\over4}R_{\mu \nu \sigma \tau}\psi^{\mu}\psi^{\nu}\psi^{* \sigma}\psi^{* 
\tau}.
\eea
The gradient flow equation of (6) is
\begin{equation}
{dx^{\mu} \over dt}=\pm g^{\mu \nu}{\partial h \over \partial x^{\nu}}.
\end{equation}
An instanton solution satisfies (7) and connects critical points.

Let $A=(a_{ij})$ be a group element of $SO(n)$.  We introduce the generalized
Euler angles ${x^{\mu}}$ as [4]
\be
A=e^{x^{n(n-1) \over 2}E_{12}} \ldots e^{x^{6}E_{12}}e^{x^{5}E_{23}}
 e^{x^{4}E_{34}}e^{x^{3}E_{12}} e^{x^{2}E_{23}}e^{x^{1}E_{12}},
\ee
where $E_{ij}$ represents a fundamental generator of a rotation in the 
$(i,j)$ plane.
The $SO(n)$ invariant metric is given by
\be
g_{\mu \nu}={1 \over 2} {\rm tr}{\partial A^{t} \over \partial x^{\mu}}
{\partial A \over \partial x^{\nu}}.
\ee
\section{Classical vacuums}
For $SO(n)$, the following is a Morse function [5]
\be
h=\sum _{i=1 }^{n} c_{i}a_{ii} ,\qquad (c_{i}>2c_{i+1}>0),
\ee
and critical points $P^{(l)}$ are
\be
P^{(l)}={\rm diag}(\epsilon _{1}, \epsilon _{2}, \ldots ,
\epsilon _{n}), \qquad (\epsilon _{i}=\pm 1, \prod _{i} \epsilon _{i}=1).
\ee
Around the critical points, $h$ is expanded as
\be
h=\sum _{i} \epsilon _{i}c_{i} +\sum _{i<j}(\lambda _{ij}\xi_{ij}^{2}+
\mu _{ij}\eta _{ij}^{2}),
\ee
where
\be
\lambda _{ij}=-{\frac{\epsilon _{j}-\epsilon _{i}} 4}(c_{j}-c_{i}),
\qquad \xi_{ij}  = a_{ij}+a_{ji},             
\ee 
\be
\mu _{ij} =-{\frac{\epsilon _{j}+\epsilon _{i}} 4}(c_{j}+c_{i}),
\qquad \eta _{ij}=a_{ij}-a_{ji}.          
\ee
Approximate vacuums are identified from  negative eigenvalues of the Hessian
matrices. From (12) we find
\be
|l>=\prod _{i < j} \prod _{\epsilon _{i}=\epsilon _{j}=1}{\hat \psi}_{\eta 
_{ij}}^{*}
\prod _{\epsilon _{i} >\epsilon _{j}}{\hat \psi}_{\xi _{ij}}^{*}|0>,
\ee
where $l$ represents the number of the excited fermions.  This state 
corresponds
to an $l$-form.

We study quantum effects between adjacent classical vacuums to select
true vacuums. According to Witten [1] the following is valid
\be
<l+1|d_{h}|l>= \sum_{\gamma} n_{\gamma} e^{-(h(P^{(l+1)})-h(P^{(l)}))},
\end{equation}
where $n_{\gamma}$ is an integer assigned for each instanton path 
$\gamma$. If a state $|l>$ does not couple with any adjacent classical
vacuums, that is if  $d _{h}|l>=<l|d _{h}=0$ , $|l>$ is a true
vacuum.

\section{Reflection Symmetries}
For $SO(n)$ the Morse function $h$ (10) is given by combinations of the 
trigonometric
functions of the Euler angles $x^{\mu}$.
Under some combinations of the reflection transformation $x^{\mu} \rightarrow
-x^{\mu}$, $h$ is invariant. These transformations are generated by $(n-1)$
transformations. The transformation that changes the sings of the off 
diagonal matrix elements with an index $i$ is one of such transformations;
\be
a _{ij},a_{ji} \rightarrow -a _{ij},-a_{ji} \quad (i: {\rm fixed}, j \ne i, 
i=2 \sim n).
\ee
We denote a supersymmetric extension of this transformation as [i]:
\be
[i]: \{x^{\mu},{\hat\psi}^{*\mu},{\hat \psi}_{\mu} \}_{i}
\rightarrow \{-x^{\mu},-{\hat\psi}^{*\mu},-{\hat \psi}_{\mu} \}_{i},
\ee
where $\{\quad  \}_{i}$ means suitable indices $\mu$ should be chosen to 
realize (17).

Under the transformation (17), ${\rm tr} A^{t}A$  is invariant and $g_{\mu 
\nu}$
and $g^{\mu \nu}$ reverse the signs as $x^{\mu}x^{\nu}$. Subsequentry, 
$\nabla_{\mu}$
has the same transformation properties as $\partial _{\mu}$. Thus, 
$d=\hat{\psi}^{* \mu} \nabla _{\mu}$ and  
$d^{\dagger}=g^{-{1 \over 2}}\nabla _{\mu} g^{1 \over 2}g^{\mu 
\nu}\hat{\psi}_{
\nu}$
are invariant under (18). Accordingly,
$d_{h},d_{h}^{\dagger}$ and $\hat H$ are also invariant.

By the transposition of the matrix $A$, we obtain one more invariant 
transformation. This transformation is represented by exchanges between some
pairs of the Euler angles. We call its supersymmetric extension $[t]$:
\be
[t]: \{x^{\mu},{\hat\psi}^{*\mu},{\hat \psi}_{\mu} \}_{t}
\leftrightarrow \{x^{\nu},{\hat\psi}^{*\nu},{\hat \psi}_{\nu} \}_{t}.
\ee
To be precise some combinations of the transformations  [i] may be 
added to (19) to represent $A \rightarrow A^{t}$. Under the transformation
[t], the metric (9) transforms as $g_{\mu \mu} \leftrightarrow g_{\nu \nu}$,
$g_{\mu \nu} \leftrightarrow g_{\mu \nu}$, 
$g_{\mu \lambda} \leftrightarrow g_{\lambda \nu}$ for a pair of indices
$\mu$ and $\nu$ in $\{\quad \}_{t}$ and $\lambda$ that does not belong to
  $\{\quad \}_{t}$ 
;as for the covariant derivatives we see
$\nabla _{\mu} \leftrightarrow \nabla _{\nu}$ and $\nabla _{\lambda} 
\rightarrow \nabla _{\lambda}$.  Thus $d_{h},d_{h}^{\dagger}$ and $\hat{H}$ 
are invariant
under $[t]$.

The above $n$ transformations generate $2^{n}-1$ symmetry transformations.
Under the transformation [i](18), ${\hat \psi}_{\eta _{lm}}^{*}$ and
${\hat \psi}_{\xi _{lm}}^{*}$ reverse the sings if $l=i$ or $m=i$. Under [t], 
${\hat \psi}_{\eta _{lm}}^{*}$ reverse the sings. Thus the classical vacuums 
have
definite parities under the symmetry transformations. If the parities
of the classical vacuums $|l>$ and $|l+1>$ are different for one of the
symmetry  transformations  the matrix element $<l+1|d_{h}|l>$ vanishes.

\section{Instanton Effects and True Vacuums}

For a couple of adjacent classical vacuums there are a pair of instanton 
paths:
\be
\left(
\begin{array}{cccccccc}
\epsilon _{1} & & & & & & &  \nonumber \\
            & \ddots & & & & & & \nonumber \\
            &   & \epsilon _{i-1} & & & & & \nonumber  \\
            &   &  & \cos \theta & \mp \sin \theta & & & \nonumber \\
    &   &  & \pm \epsilon \sin \theta & \epsilon \cos \theta & & & \nonumber 
\\
            &   &  &   &    & \epsilon _{i+2}   &    & \nonumber \\
            &   &  &   &    &    & \ddots   & \nonumber \\
            &   &  &   &    &    &    &  \epsilon _{n}  \\
\end{array}
\right),\\
\ee
where $\epsilon=\pm 1$ . These two paths are invariant or exchanged each
other by the symmetry transformations.   
These two paths will give a pair of instanton solutions with the 
Euler angles except for
$\theta$ are constants.   For the paths (20), $g_{\theta \theta}=
g^{\theta \theta}=1$, and the gradient flow equation (7) for $\theta$ will be
\be
{d\theta \over dt} =-(c_{i} \pm c_{i+1})\sin \theta.
\ee
Equation (21) has the general instanton-type solution
\be
\cos \theta = \tanh ((c_{i} \pm c_{i+1})t+\alpha),
\ee
with $\alpha$  a constant.

An instanton solution causes a non-zero  instanton effect between the 
corresponding couple of classical vacuums. However, a pair of instanton 
effects
can cancel each other. It is crucial to determine whether a couple of 
instanton effects cancel each other or not. If the matrix element 
$<l+1|d_{h}|l>$ vanishes
by the symmetries, we see the two instanton effects cancel each other.
If the matrix element does not vanishes by the symmetries, it will be 
plausible
to say the two instanton effects add up. Thus we can select true vacuums by
the reflection symmetries. The number of selected vacuums will be in 
agreement
with the de Rham cohomology of $SO(n)$. In the next section  we discuss
$SO(5)$ as an example. We see our selection rule works well.

\section{An Example: $SO(5)$}
There are 16 classical vacuums
\begin{eqnarray}
|0>, (-1,-1,-1,-1,1)&;& \quad |1>,  (-1,-1,-1,1,-1)\quad;\quad|2>,
  (-1,-1,1,-1,-1); \nonumber \\ 
|3A>,\quad  (-1,-1,1,1,1)&;&\quad|3B>,  (-1,1,-1,-1,-1);\quad
|4A>,  (-1,1,-1,1,1);\nonumber \\
|4B>,  (1,-1,1,-1,-1)&;& \quad
|5A>,  (-1,1,1,-1,-1)\quad;\quad|5B>,  (1,-1,-1,1,1);\nonumber \\
|6A>,\quad  (-1,1,1,1,-1)&;&\quad|6B>,  (1,-1,1,-1,1)\qquad;\quad 
|7A>,  (1,-1,1,1,-1); \nonumber \\
|7B> ,\quad(1,1,-1,-1,1)&;&\quad 
|8>, (1,1,-1,1,-1)\qquad;\quad|9>,  (1,1,1,-1,-1); \nonumber \\
|10>,\quad \qquad (1,1,1,1,1)&,&
\end{eqnarray}
where for example $|4A> \sim {\hat \psi}^{*}_{ \xi _{23}}{\hat \psi}^{*}_{ 
\eta _{24}}
 {\hat \psi}^{*}_{ \eta _{24}}{\hat \psi}^{*}_{ \eta _{45}}|0>$.
In Table 1. we denote their parities. From the selection rule we see
$|0>, |3A>, |7B>$ and $|10>$ are true vacuums. The other classical
vacuums will cease to be vacuums by quantum effects. This result is 
in agreement with  $H^{*}(SO(5)) \cong \wedge (x_{3},
x_{7}).$

\begin{center}
\vspace{5mm}
\begin{tabular}{|c|c|c|c|c|c|c|c|c|c|c|c|c|c|c|c|c|} \hline
    & 0& 1 &  2 & 3A &   3B & 4A &  4B
 & 5A &5B &  6A & 6B &  7A & 7B & 8& 9 & 10 \\ \hline
     &  &  &  &  &  &  &  &  &  &  &  &  &  &  &  &   \\  
$[2]$  & e & e & e & e  & o & o & o & o & o & o  & o & o & e  & e & e
 & e \\ \hline
     &  &  &  &  &  &  &  &  &  &  &  &  &  &  &  &\\  
$[3]$ & e & e & e & e  & o & o & o & o & o & o & o & o & e & e & e &
e \\ \hline
     &  &  &  &  &  &  &  &  &  &  &  &  &  &  &  & \\ 
$[4]$ & e & o & o & e & o & e & o & e & e & o & e & o & e & o & o & 
 e \\ \hline
     &  &  &  & &  &  &  &  &  &  &  &  &  &  &  &\\ 
$[5]$ & e & o & o & e & o & e & o & e & e & o & e & o & e & o & o & 
  e \\ \hline
     &  &  &  & &  &  &  &  &  &  &  &  &  &  &  &\\
$[t]$ & e & e & e & o & e & o & e & o & o & o & o & o & o & o & o &
 e \\ \hline
\end{tabular}
\\ 
\end{center}
Table.1.
\\
\\
{\bf Acknowledement} \\
The author thanks Dr.Yasui for introducing him to this subject.
\\
\\
{\bf References}  \\
1. Witten,E.:{\it J.Diff.Geom.}{\bf 17}(1982),661-692. \\
2. Hirokane,T.,Miyajima,M. and Yasui,Y.:{\it J.Math.Phys.}{\bf 
34}(1993),2789-2806. \\
3.Oshima, K.: Instanton Effects and Witten Complex in Supersymmetric
Quantum Mechanics on $SO(4)$,Preprint   hep-th/9508171. \\
4.B\"{o}hm, M. and Junker, G.:{\it J.Math.Phys.}{\bf 28}(1987),1978-1994. \\
5.Yokota,I.:{\it Manifold and Morse Theory(in Japanese)},Gendai 
Suugakusha,Kyoto,1989,pp153-162 . \\

\end{document}